\documentclass{kapproc} 

\usepackage[dvips]{graphicx}
\usepackage{t1enc}
\usepackage{procps}
\usepackage{epsf}
\usepackage{amssymb} 
\usepackage{latexsym}
\def\ltsim{~\rlap{\lower -0.5ex\hbox{$<$}}{\lower 0.5ex\hbox{$\sim\,$}}}
\upperandlowercase
\kluwerbib 
\begin{document}
\articletitle[The outer disks of galaxies]{The outer disks of galaxies: \\
``To be or not to be truncated?''}
\chaptitlerunninghead{The outer disks of galaxies}
\author{Michael Pohlen\altaffilmark{1}
and Ignacio Trujillo\altaffilmark{2}}
\altaffiltext{1}{Kapteyn Astronomical Institute, University of Groningen, 
The Netherlands}
\altaffiltext{2}{Max--Planck--Institut f\"ur Astronomie, Heidelberg, Germany}
\begin{abstract}
We have in recent years come to view the outer parts of galaxies as 
having vital clues about their formation and evolution. Here, we would like 
to briefly present our results from a complete sample of nearby, late-type, 
spiral galaxies, using data from the SDSS survey, especially focused on 
the stellar light distribution in the outer disk. 
Our study shows that only the minority of late-type galaxies show a 
classical, exponential Freeman Type\,I profile down to the 
noise limit, whereas the majority exhibit either downbending
(stellar truncation as introduced 1979 by Piet van der Kruit) or 
upbending profiles. 
\end{abstract}
\section{Historical introduction}
\subsubsection*{Why study outer disks?}
The structure of galactic disks is of fundamental importance for
observationally addressing the formation and evolution of spiral 
galaxies. 
Especially the {\it outer edges} of disk galaxies are of interest,
since substructure, the so called fossil evidence, is expected 
to be imprinted by the galaxy formation process. 
Recent formation and evolution scenarios suggest for example  
that galaxies continue to grow from the accretion and tidal 
disruption of satellite companions.  
%
%
\subsubsection*{What is their shape?}
The general shape of the surface brightness profile of galactic 
disks is currently one of our favourite paradigms, namely
a pure exponential disk. 
Its success is based solely on empirical evidence (albeit dating 
back now nearly 50 years) and has in fact never been fully physically 
motivated.
\begin{quote}
In view of the great variety of structures among spirals ... there is 
good evidence, however, that at least in ordinary spirals the smoothed 
radial luminosity distribution is approximately exponential in the 
outer parts (de Vaucouleurs 1959)
\end{quote}
Eleven years later the general nature of disks was finally settled by 
Ken Freeman in his 1970 paper.
\begin{quote}
Almost every disklike galaxy with measured $I(R)$ shows an exponential 
disk ... and its origin is certainly a significant cosmogonic problem.
(Freeman 1970) 
\end{quote}
However, only nine years later Piet van den Kruit indicated 
that the exponential nature does not hold to infinity, but 
\begin{quote}
[..] at the edges of the disk the decrease in apparent surface brightness 
is exceedingly steep. This sharp drop implies that galaxies do not retain 
their exponential light distribution to such faint levels.  
(van der Kruit 1979)
\end{quote}
This marked the detection of truncations at a safe distance of 
$\sim\,4.5$ times the radial scalelength, so the paradigm of the 
exponential disk was not in real danger at the time.        
%
%
\subsubsection*{Where are they truncated?}
After van der Kruit and Searle's seminal papers about the structure 
of galactic disks (e.g. van der Kruit \& Searle 1981), the matter of 
imperfectly exponential disks was rather ignored for some more ten years, 
until Barteldrees \& Dettmar~(1994) confirmed their existence using for 
the first time modern CCD equipment, but placed the cut-off closer to 
the center, for some galaxies at a disturbing close $<\,3$ times the 
scalelength.    
\newline
Finally starting from the year 2000 several groups 
(e.g. Pohlen et al.~2000, de Grijs et al.~2001, Florido et al.~2001,
Kregel et al.~2002, Pohlen et al.~2002, or Erwin et al.~2005a)
followed up the question of where the disks are truncated, how the shape 
of the profile in the very outer parts looks like, and if all disks
have a truncation. For a recent review see Pohlen et al.~(2004).
\newline
To keep a long story short our answer is that we know now that not 
all galaxies are truncated, but those which are, are now believed 
to be truncated 'early' (at $\sim 3 h$), often abruptly (but not 
completely) and the profiles are best described as a broken exponential 
and so probably better called {\it breaks} than {\it cut-offs}.  
The prototypical break is recently observed for M\,33 by Ferguson 
(this volume) using the star-count method as an independent 
approach compared to the so far purely surface photometric measurements. 
The currently favoured origin for these breaks are global star-formation 
thresholds as described by Martin \& Kennicutt (2001) or Schaye (2004),
although this does not explain the origin of the material beyond 
the break (see Pohlen et al.~2004). 
%
\subsubsection*{What next?}
What we still need is a complete census of the outer 
disk structure in the local universe extending the work done by 
Courteau (1996) and de Jong (1996) including a detailed 
discussion of the light profile in the outer region to answer
our famous question: ``To be or not to be (truncated)?''.  
This is now done and will be presented briefly here and later 
in detail in Pohlen \& Trujillo (2005, in prep.) for late-type 
galaxies and in Erwin et al.~(2005b, in prep.) for early-type
galaxies.
%
\section{Sample}
We used the LEDA online catalogue (the richest, most complete and 
up-to-date catalogue with homogeneous parameters of galaxies for 
the largest available sample) to select our initial galaxy sample
using the following selection criteria: $2.99 < T < 8.49$ (Sb-Sdm), 
$\log r_{25} < 0.301$, $v_{\rm vir} < 3250$ km/s,
$|b2| > 20^{\circ}$, and $M_{\rm abs}  < -18.4$ B-mag.
This leaves us with an unbiased, volume limited sample of late-type, 
face-on ($i\ltsim60\deg$), nearby (local) disk galaxies (avoiding the 
Milky Way). Our sample complements the CCD imaging 
sample of $\sim\,65$ early-type SB0-SBb galaxies by Erwin et al.~(2005b, 
in prep.).
The actual data we use for our present study come from the 
Sloan Digital Sky Survey (SDSS, data release 2) providing images 
of $\sim 15\%$ (98/655) of the galaxies in our original LEDA sample. 
%
\section{Results}
\subsubsection*{SDSS profiles:}
To convince the reader, and first of all ourselves, that SDSS images with
a rather short exposure time of only $\sim\,60$s are 
indeed deep enough to trace the outer disk we compared for three galaxies 
SDSS images with deep surface photometry. The deep data is presented in 
Pohlen et al.~(2002) and was obtained at the CAHA 2.2m telescope
using CAFOS. The total exposure time of these images is about 3 hours
reaching reliably down to $\mu_{\rm lim}\,=\,27.2$ R-mag/sqarcsec. 
As shown in Fig.1 (for two of the galaxies) the agreement is astonishingly 
good, allowing us to safely use SDSS images to study the profile clearly 
beyond the break radius. 
\begin{figure}[h]
\hspace*{-1.0cm}
\includegraphics[angle=270,width=6.8cm]{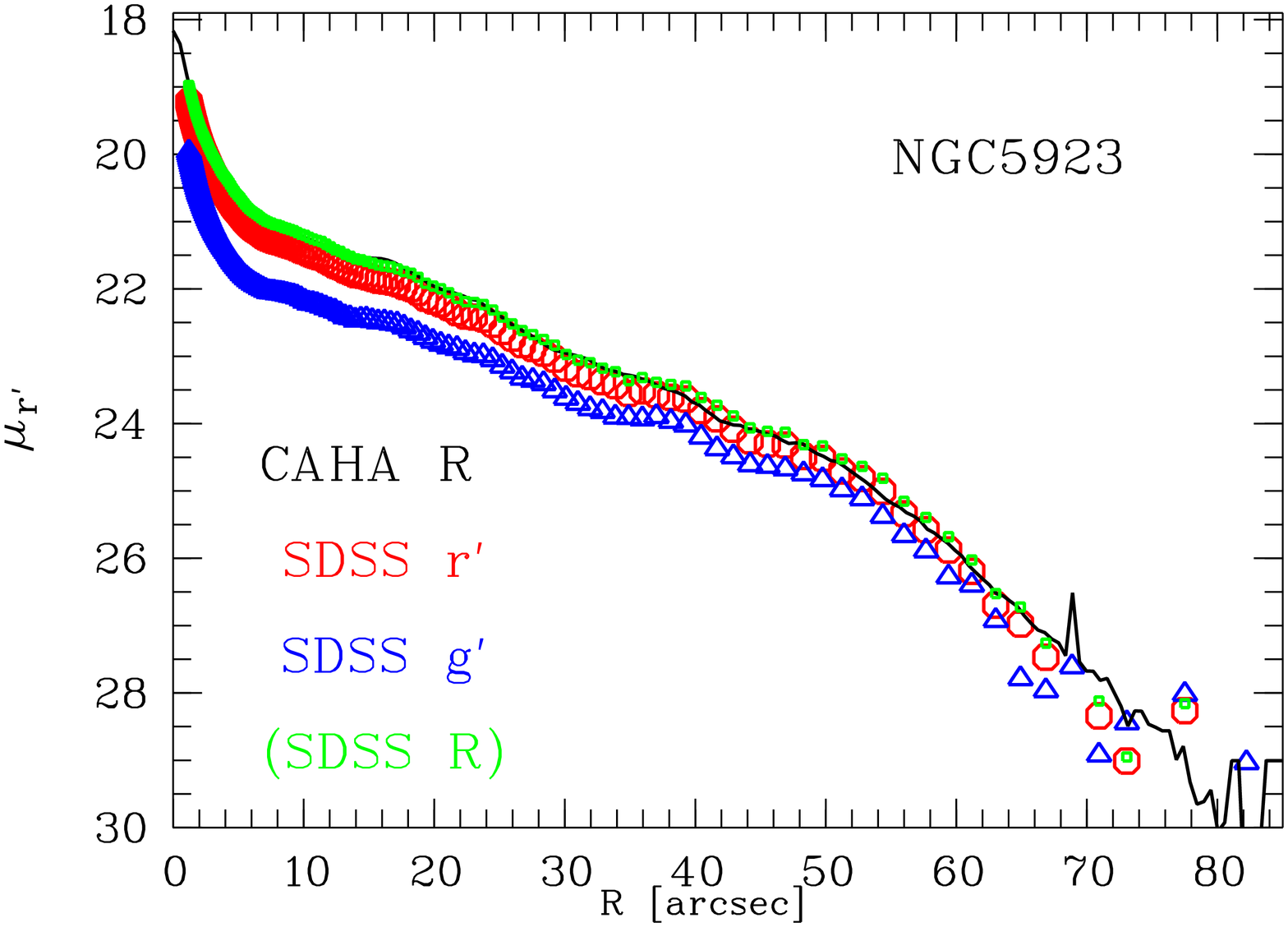}
\includegraphics[angle=270,width=6.8cm]{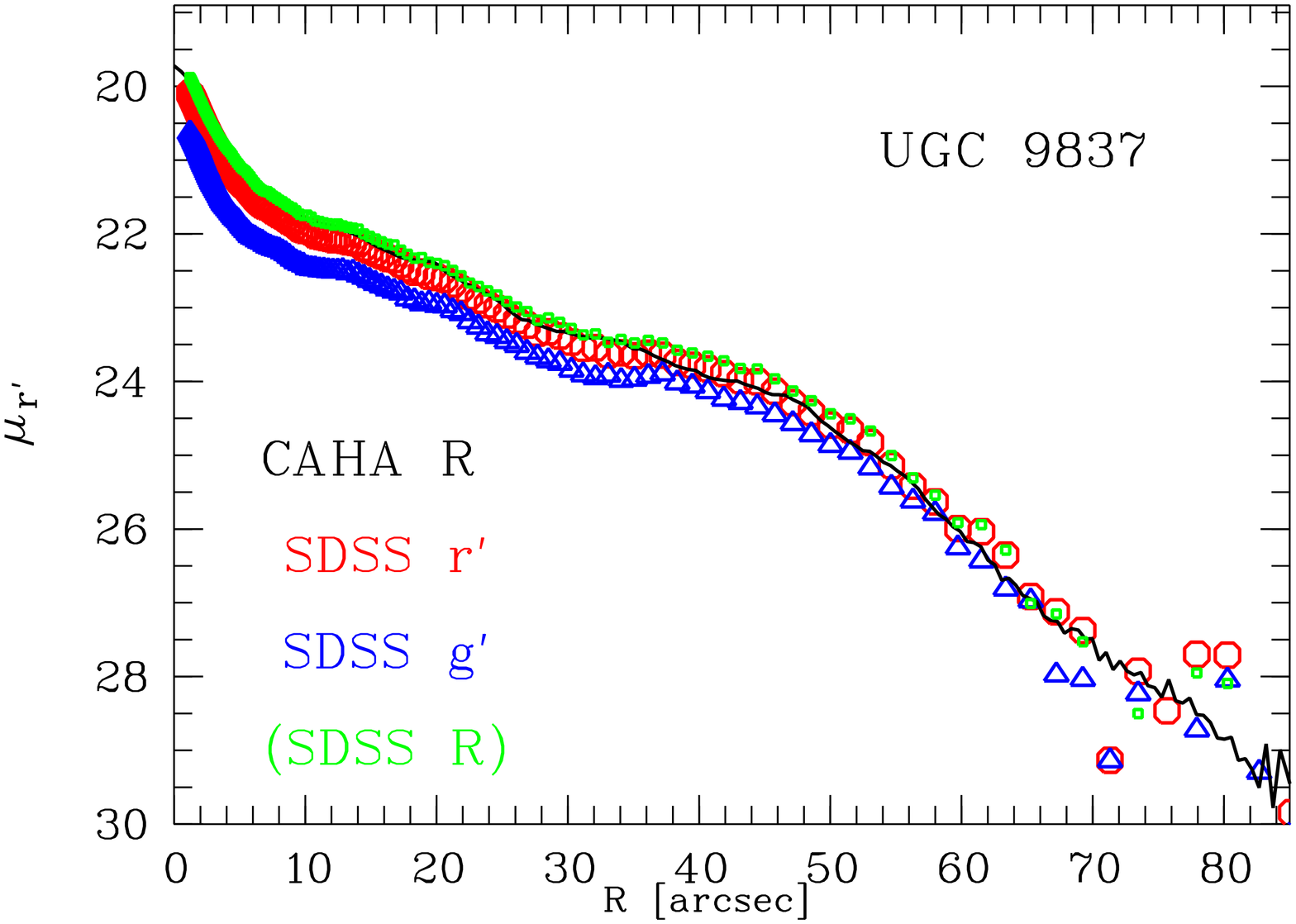}
\caption{Comparison of the azimuthally averaged, radial surface 
brightness profiles from SDSS images with much deeper imaging 
by Pohlen et al.~(2002). The small, green squares are produced 
by using the transformation to convert SDSS g' and r' into standard 
Johnson R band following Smith et al.~(2002).}
\end{figure}
%
%
\subsubsection*{Classification:}
We classify each galaxy profile by eye according to the following 
nomenclature.
If there is no indication for any obvious break in the profile the 
galaxy is classified as Type\,I following Freeman (1970). 
In the same sense, galaxies showing a profile better described as 
a broken exponential with a clear break and a {\it downbending}, 
steeper outer region are defined here as Type II:
\begin{quote}
Type\,II has $I(R) < I_{0} \exp(-\alpha R)$ in an interval 
$R_1\,<\,R\,<\,R_2$ not far from the center. (Freeman 1970)
\end{quote}
This definition includes the class of truncated galaxies shown 
by Pohlen et al.~(2002). 
Although ``not far from the center'' may suggest to exclude 
truncated galaxies (where the break is at several radial 
scalelengths) from this class, thus following quote allows 
us to generalise Ken's definition:   
\begin{quote}
It is worth pointing out that for the $[$...$]$ galaxies in which the 
Type\,II characteristic is most prominent $[$e.g. NGC 7793$]$, the 
exponential disk begins outside the main region of the spiral-arm 
activity. (Freeman 1970)
\end{quote}
Finally, following Erwin et al.~(2005a), galaxies showing a 
broken exponential profile with a break, but followed by an 
{\it upbending}, shallower outer region, are called Type\,III.
Three proto-typical cases are shown in Fig.2. 
\begin{figure}[h]
\hspace*{-1.0cm}
\includegraphics[angle=270,width=6.8cm]{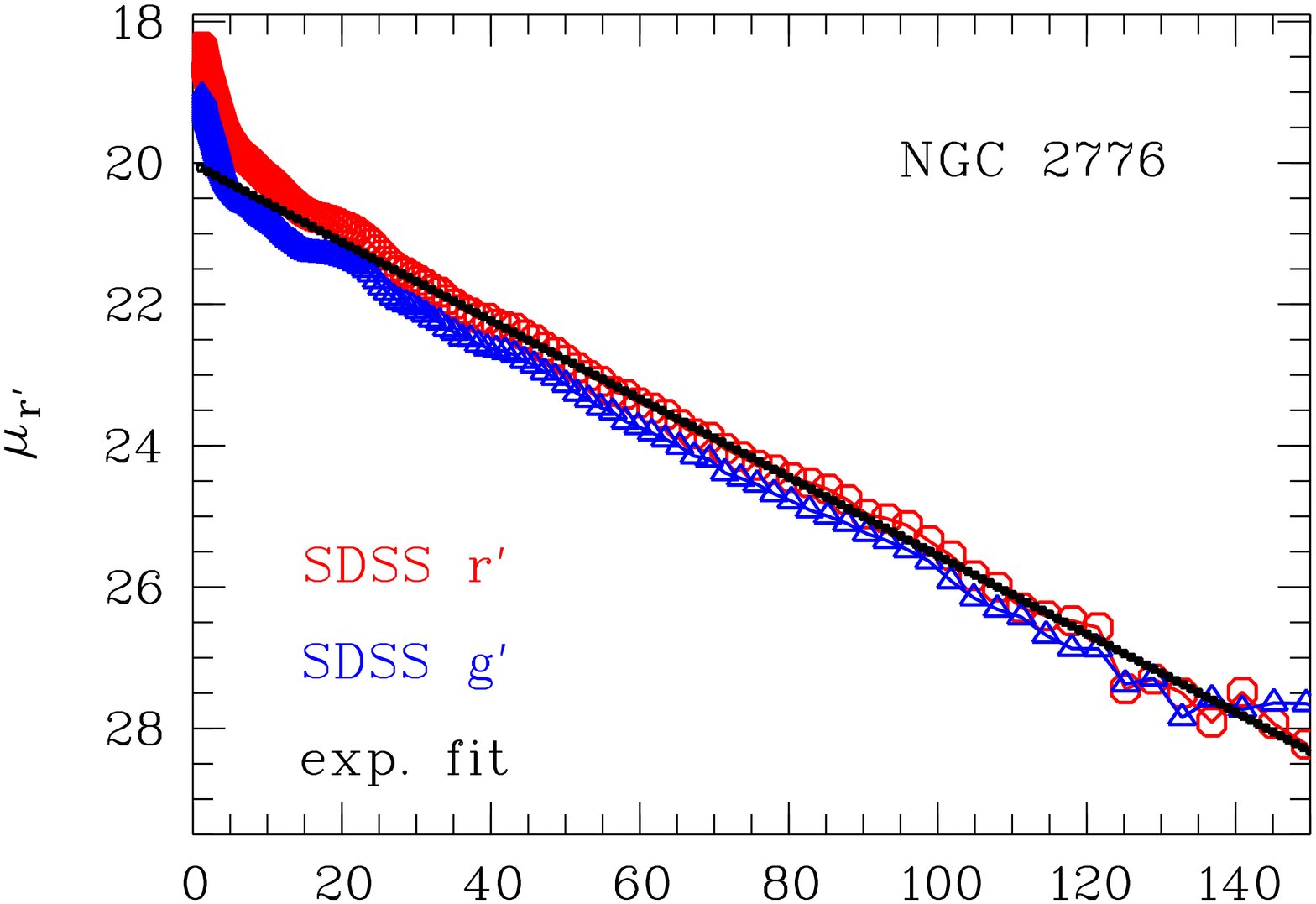}
\includegraphics[angle=270,width=6.3cm]{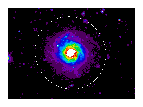}
\hspace*{-1.0cm}
\includegraphics[angle=270,width=6.8cm]{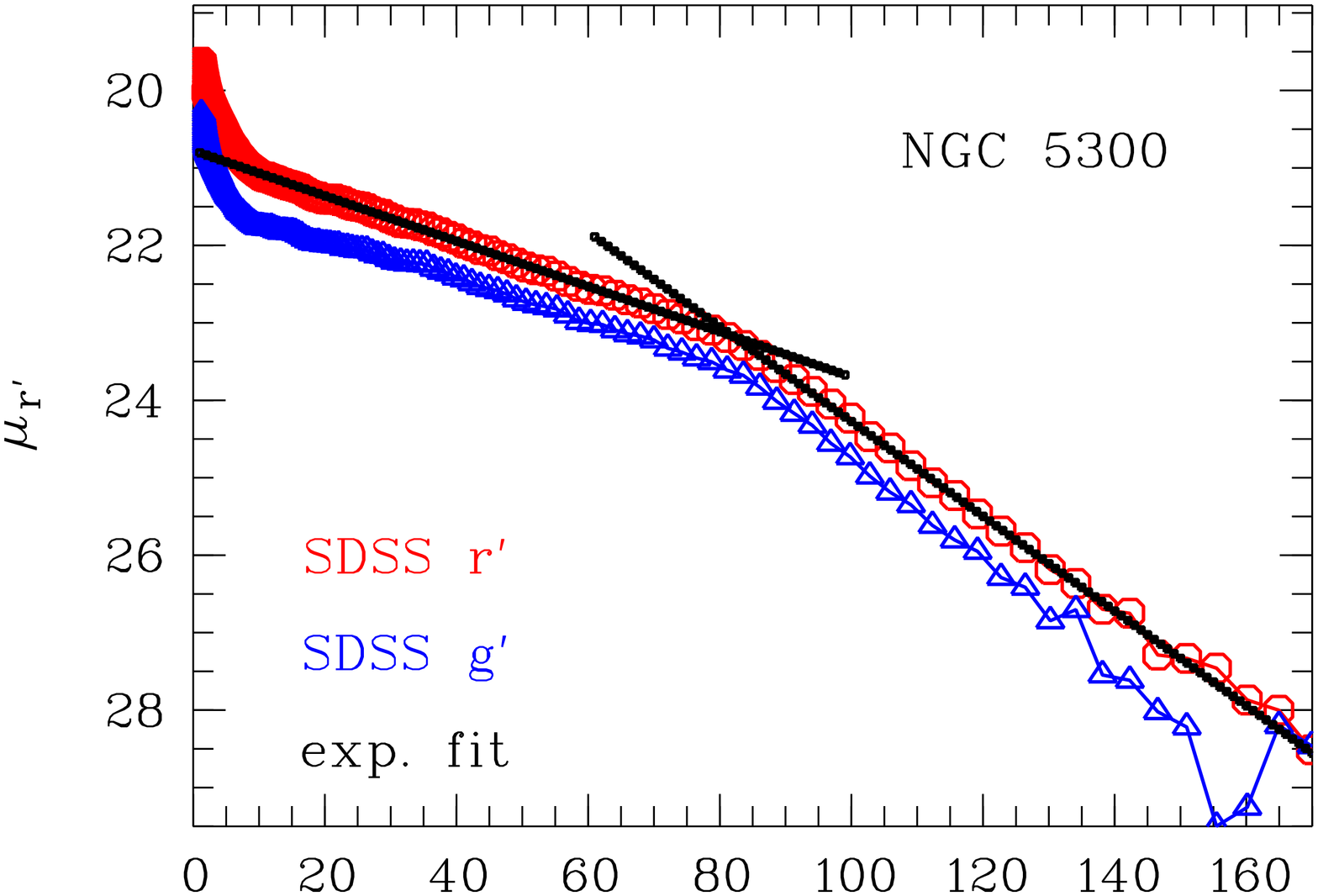}
\includegraphics[angle=270,width=6.1cm]{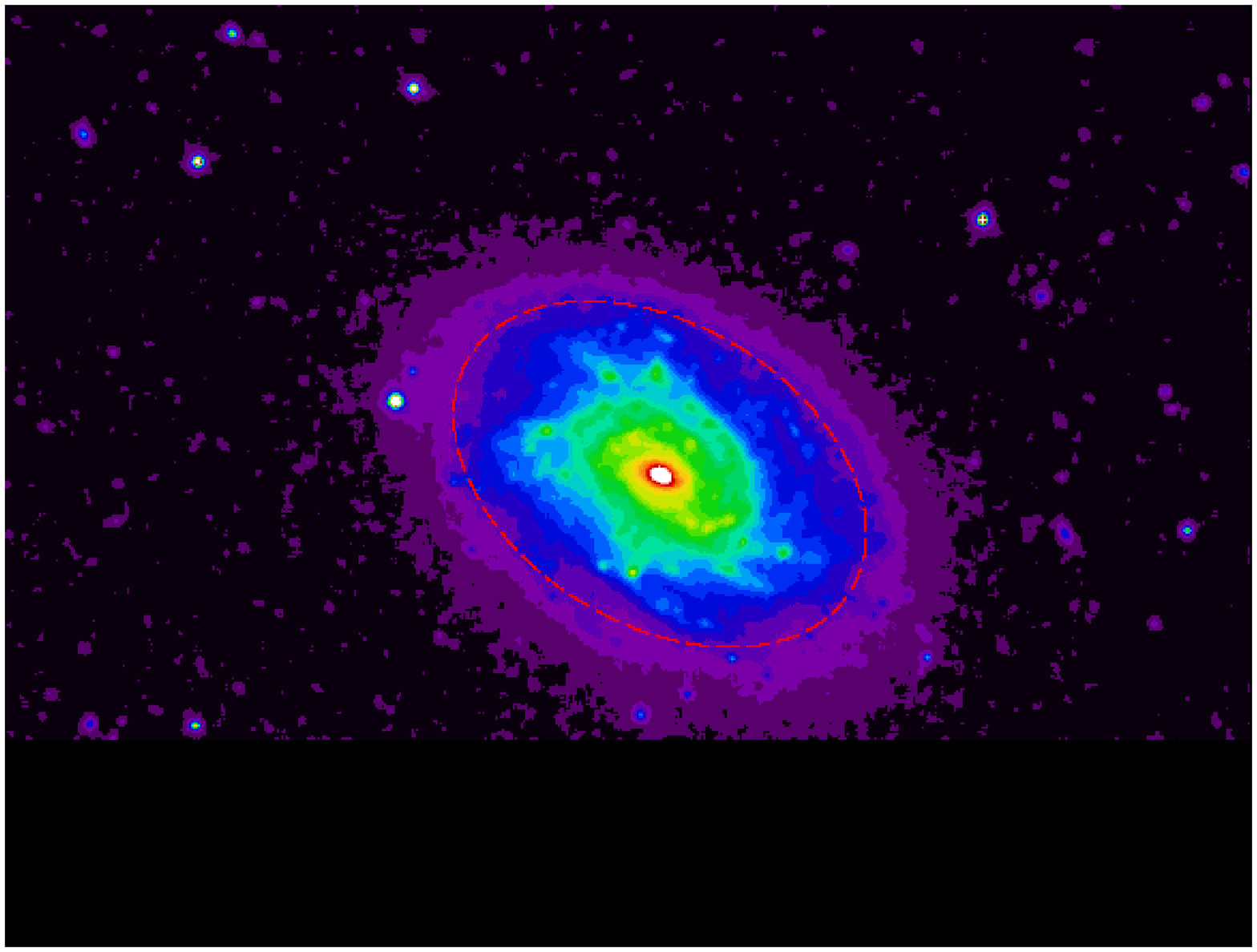}
\hspace*{-1.0cm}
\includegraphics[angle=270,width=6.8cm]{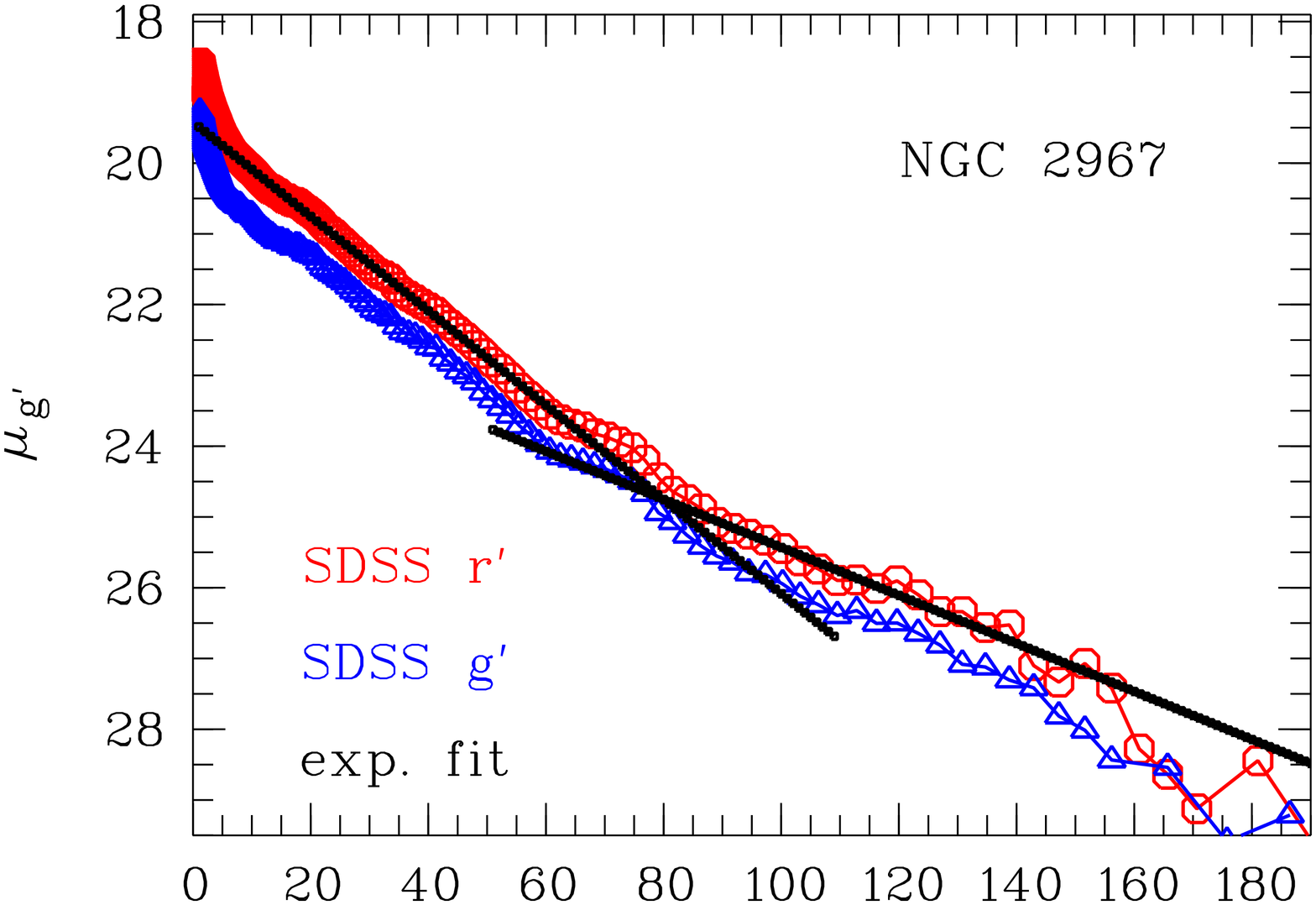}
\includegraphics[angle=270,width=6.3cm]{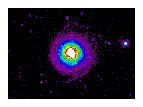}
\caption{The three main disk types: Type\,I, Type\,CT, and 
Type\,III {\sl (from top to bottom)}.
Left column: Azimuthally averaged, radial (in units of $['']$) 
SDSS surface brightness profiles in the g' and r' band overlayed 
by r' band exponential fits to the individual regions: single disk; inner 
and outer disk. Right column: r' band images with the break radius 
marked as a red ellipse. The white ellipse for the first Type\,I 
galaxy corresponds to roughly the noise limit at $\sim\,140''$.}
\end{figure}
%
\subsubsection*{Frequencies and parameter distribution:}
The Type\,II class is split into several sub classes (discussed
in detail by Pohlen \& Trujillo, 2005, in prep.) but the majority, about 
$32.9\% \pm 5.1\%$ of the total sample, are what we call now {\it 
classical truncations} 
(Type\,CT), associated with those having a star formation threshold origin.  
An equal amount of galaxies (also $32.9\%$) are in fact {\it anti-truncated} 
and fall into the Type\,III class. 
Finally, only the minority ($15.3\% \pm 3.9\%$) of galaxies is barely 
consistent with being pure exponential disks of Type\,I.
\newline
We find indications for a trend with Hubble Type: Early-type galaxies
(Sb-Sc) are more commonly Type\,III while Type\,CTs seem to
be significantly more frequent in later types. 
Counting the neighbouring galaxies around each galaxy reveals the fact 
that Type\,III galaxies statistically prefer a high density, Type\,CT 
galaxies a low density environment. However, this relation is far 
from being clear-cut. 
\newline
The break radius for galaxies with classical truncations, the ones 
discovered by Piet van der Kruit in 1979, happens in the r' band at 
$R_{\rm break}\,=\,2.5\,\pm\,0.6\  h_{\rm inner}$ ranging between 
1.4 and 4.2.
Since we can follow the profile of the Type\,I galaxies down to 
$\sim 6-8$ times their scalelength, they seem to be genuinely
untruncated galaxies.
\newline
While the scalelength varies with filter (being larger in g' compared 
to r') as known before, we do not find a systematic difference for the 
break radius.
We do, however, find that the distribution of the surface brightness 
at the break radius $\mu_{\rm break}$ for galaxies with a classical 
truncation is peaked around a mean value. Together with 
the absence of a relation between $R_{\rm break}/h_{\rm inner}$ and 
mass (rotational velocity), this favours a star formation threshold
scenario for its origin.
%
\section{Stellar Disks truncations at high-z}
P\'erez (2004, and this volume) showed that it is possible 
to detect truncations even out to high redshift ($z\sim\,1$).
So we carefully defined a complete sample of high redshift galaxies 
(Trujillo \& Pohlen, 2005) using the ACS data of the Hubble Ultra 
Deep Field.
From the final sample of 36 galaxies, 21 show truncations. 
Now, using the position of the truncation as a direct estimator 
of the size of the stellar disk it becomes possible to outright
observe inside-out growth of galactic disks comparing the ACS 
to our local SDSS sample.
The results suggest that the radial position of the truncation
has increased with cosmic time by $\sim\,1-3$ kpc in the last 
$\sim\,8$ Gyr indicating a small to moderate ($\sim25\%$) inside-out 
growth of the disk galaxies since $z\sim\,1$ (see Trujillo \& Pohlen, 2005).
\begin{acknowledgments}
M.P. would like to thank Peter Erwin and John Beckman for their 
stimulating discussions and useful suggestions during this work. 
Part of this work was supported by a Marie Curie Intra-European 
Fellowship within the 6th European Community Framework Programme.
\end{acknowledgments}
%

%
\end{document}